\documentclass[letterpaper,11pt]{article}

\usepackage[]{hyperref}
\usepackage{amsmath,amssymb,graphicx,physics,siunitx}
\usepackage{amsthm,bm,dcolumn}
\usepackage{graphicx}
\usepackage{geometry}
\usepackage[square,sort,comma,numbers]{natbib}

\title{Generalized Aharonov-Bohm Effect} 

\author{Shan Gao \\
Research Center for Philosophy of Science and Technology, \\
Shanxi University, Taiyuan 030006, P. R. China \\
E-mail: \href{mailto:gaoshan2017@sxu.edu.cn}{gaoshan2017@sxu.edu.cn}}

\begin{document}

\maketitle

\begin{abstract}
The Aharonov-Bohm (AB) effect highlights the fundamental role of electromagnetic potentials in quantum mechanics, manifesting as a phase shift for a charged particle in field-free regions. While well-established for static magnetic fluxes, the effect’s behavior under time-varying fluxes remains an open and debated question. Employing the WKB method, we derive the AB phase shift for a time-dependent magnetic vector potential, demonstrating that for circular paths in the quasistatic regime, it is proportional to the time-averaged enclosed magnetic flux, \(\Delta \phi_{\rm AB} = \frac{1}{T} \int_0^T e \Phi(t) \, dt\), with the total phase shift, including kinetic contributions, equaling \(e \Phi(0)\). For non-circular paths, the phase shift depends on both the flux history and path geometry, revealing the effect’s hybrid nature involving gauge potentials and induced electric fields. We verify the consistency of our gauge choice with Maxwell’s equations and discuss the implications for local versus nonlocal interpretations of the AB effect. We also generalize the results to scenarios with nonzero external magnetic fields, where the enclosed flux is through the actual electron paths, and for circular paths of radius $R$, the AB phase shift is also proportional to the time average of the enclosed flux \(\Phi_{\rm enc}(R,t)\), with the total phase shift depending only on the initial enclosed flux \(e \Phi_{\rm enc}(R,0)\); for general non-circular paths, the external magnetic field affects trajectories and phase accumulation through the Lorentz force, leading to additional path dependence. These findings clarify the role of gauge-dependent potentials and induced fields in the generalized AB effect, offering new theoretical insights and potential applications in quantum technologies.
\end{abstract}

Keywords: quantum mechanics; magnetic vector potential; time-varying magnetic flux; WKB method; generalized Aharonov-Bohm effect; phase accumulation

\section{Introduction}

The Aharonov-Bohm (AB) effect, first predicted by Werner Ehrenberg and Raymond E. Siday in their 1949 paper on electron optics but largely overlooked at the time, was independently rediscovered and given a broader theoretical framework in quantum mechanics by Yakir Aharonov and David Bohm in 1959, making it one of the most intriguing phenomena that challenges classical intuitions about physical interactions \cite{Ehrenberg1949, Aharonov1959, Aharonov1961, Hiley2013, Aharonov2025}. In its canonical magnetic form, a charged particle, such as an electron, acquires a phase shift when passing around a solenoid enclosing a magnetic flux, despite experiencing no local electromagnetic force. This phase shift, observable as a shift in the interference pattern, arises from the magnetic vector potential and highlights the nonlocality of quantum mechanics and the fundamental role of gauge potentials. 

The static AB effect has been extensively validated through experiments, such as electron interferometry \cite{Chambers1960,Tonomura1986}, and theoretically analyzed within the framework of gauge theories \cite{Peshkin1989, Aharonov2025}. In the static case, the phase shift is directly proportional to the enclosed magnetic flux, as determined by the gauge-invariant line integral of the vector potential around a closed path. However, extending the AB effect to time-varying magnetic fluxes introduces significant theoretical and experimental challenges, with conflicting predictions in the literature \cite{Lee1992, Gaveau2011, SV2013, MS2014, BS2015, AS2016, JZWLD2017, CM2019, Wakamatsu2024, Wakamatsu2025}. Some studies argue that the phase shift remains static, determined by the initial flux value \cite{SV2013, MS2014}, while others propose a dynamic phase shift that evolves with the time-varying flux \cite{Lee1992, JZWLD2017, CM2019}. These discrepancies stem from differing treatments of the induced electric field and the electron’s dynamic motion, as well as assumptions about the quasistatic approximation \cite{Abbott1985, Templin1995, Parker2024, Zangwill2013}.

Resolving these debates is crucial for understanding the interplay between time-dependent electromagnetic potentials and quantum phase accumulation, with implications for gauge theories \cite{Gao2025}. The time-dependent AB effect also holds potential for applications in quantum technologies where precise control of quantum phases is essential. Previous analyses have often relied on gauge transformations or simplified path assumptions, which may not fully capture the dynamic effects of time-varying potentials, particularly when induced electric fields are present \cite{Wakamatsu2024, Wakamatsu2025}.

In this work, we derive the AB phase shift for time-dependent magnetic vector potentials using the WKB method, which aptly incorporates electron trajectories and induced fields. We decompose the total phase into AB and kinetic components, computing angular velocities influenced by the induced electric field. For circular paths in the quasistatic regime, the AB phase shift equals the time-averaged enclosed flux multiplied by the charge, $\Delta \phi_{\rm AB} = \frac{1}{T} \int_0^T e \Phi(t) \, dt$, with the kinetic contribution yielding a total phase of $e \Phi(0)$. For non-circular paths, we furnish a general framework where the total phase depends on the flux history $\Phi(t)$ and path geometries, emphasizing path dependence in dynamic scenarios. 

Moreover, we further generalize these results to cases with nonzero external magnetic fields, where for circular paths of radius $R$, the AB phase shift is the time average of the enclosed flux \(\Phi_{\rm enc}(R,t)\), with the total phase depending only on the initial enclosed flux \(e \Phi_{\rm enc}(R,0)\); for general non-circular paths, the external magnetic field affects trajectories and phase accumulation through the Lorentz force, leading to additional path dependence, and reducing to the quasistatic results when radiative effects are negligible. We also verify the consistency of our gauge choice in the quasistatic regime with Maxwell’s equations and discuss the implications for local versus nonlocal interpretations of the AB effect. These findings clarify the role of gauge-dependent potentials and induced fields in the generalized AB effect, offering new theoretical insights and potential applications in quantum technologies.

\section{Derivation of the Phase Shift for Time-Varying Magnetic Flux}

The Schr\"{o}dinger equation for a charged particle such as an electron in the presence of an electromagnetic potential (in units where $\hbar=c=1$) is given by:
\begin{equation}
    i\frac{\partial \psi}{\partial t} = -\frac{1}{2m} \left(\nabla - ie \mathbf{A} \right)^2 \psi + e A_0 \psi,
\end{equation}
where $e$ and $m$ are respectively the charge and mass of the electron, and $\mathbf{A}$ is the magnetic vector potential and $A_0$ is the electric scalar potential. 
When electromagnetic fields are present along the electron paths as in the time-dependent AB effect, the potentials $A_0$ and $\mathbf{A}$ cannot be gauged away to yield a free Schrödinger equation. The WKB method is necessary and appropriate, leveraging the well-localized nature of the wave packets to approximate the phase via a Hamilton–Jacobi-like equation, capturing the AB phase shift while accounting for the potentials and any field effects.

With the standard WKB ansatz $\psi=R e^{iS/\hbar}$ and taking leading order in $\hbar$, the phase change of the wave function along a trajectory $\vb r(t)$ is given by 
\begin{equation}
\Delta \phi = S(t_1)-S(t_0) = \int_{t_0}^{t_1}\left[\tfrac12 m v^2 + e\vb A(\vb r,t)\cdot {\vb v}(\vb r,t) - eA_0(\vb r,t)\right]dt.
\end{equation}
This phase change is composed of both the standard AB part $\Delta \phi_{\rm AB}$ and the kinetic part $\Delta \phi_{\rm kin}$, where 
\begin{equation}
\Delta \phi_{\rm AB} = \int_{t_0}^{t_1}\left[e\vb A(\vb r,t)\cdot {\vb v}(\vb r,t) - eA_0(\vb r,t)\right]dt
\end{equation} 
and
\begin{equation}
\Delta \phi_{\rm kin}  = \int_{t_0}^{t_1}\tfrac12 m v^2dt.
\end{equation} 
In the following, we will calculate these two parts respectively. 

Consider the magnetic AB effect. 
A beam of electrons emitted from a source is split into two parts, each going on opposite sides of a solenoid. After the beams pass by the solenoid, they are combined to interfere coherently. 
For an infinitely-long solenoid with time-dependent magnetic flux $\Phi(t)$, 
we can choose a gauge in which $A_0(\bm{r},t)=0$ and 
\begin{equation}
    \mathbf{A}(\bm{r},t) = \frac{\Phi(t)}{2\pi r} \hat{\bm{\theta}} \label{A}
\end{equation}
for the region outside the solenoid, where $\hat{\bm{\theta}}$ is a unit vector in the angular direction.\footnote{This gauge choice is widely used in studies of time-dependent AB effect \cite{Lee1992, Gaveau2011, SV2013, MS2014, BS2015, AS2016, JZWLD2017, CM2019, Wakamatsu2024, Wakamatsu2025}. But it is only a quasistatic approximation. We will discuss its consistency with Maxwell’s equations later.}  Then, we can obtain the AB phase shift:
\begin{equation}
    \Delta \phi_{\rm AB} = e \int_{L_1} \mathbf{A(r},t) \cdot d\mathbf{r}
    - e \int_{L_2} \mathbf{A(r},t) \cdot d\mathbf{r} = 
    e \oint_C \mathbf{A(r},t) \cdot d\mathbf{r},\label{AB0}
\end{equation} 
where $L_1$ and $L_2$ are the paths of the two electron beams respectively, and $C$ is the whole closed path around the solenoid. 
In the time-independent case where $\Phi(t)=\Phi_0$, this simplifies to (by Stokes’ theorem): 
\begin{equation}
    \Delta \phi_{\rm AB} = e\Phi_0.
\end{equation}
However, when $\Phi(t)$ varies with time, we must consider the motion of the electron around the solenoid in order to calculate the AB phase shift. 
Substituting (\ref{A}) in the phase shift integral (\ref{AB0}) we have: 
\begin{equation}
    \Delta \phi_{\rm AB} = e \oint_C \frac{\Phi(t)}{2\pi r} \hat{\bm{\theta}} \cdot d\mathbf{r}.
\end{equation}
Since $\hat{\bm{\theta}} \cdot d\mathbf{r}=\omega(t)rdt$, we obtain: 
\begin{equation}
    \Delta \phi_{\rm AB} = \frac{e}{2\pi} \int_0^{T}{\Phi(t)(\omega_1(t)+\omega_2(t))dt},\label{AB}
\end{equation}
where $\omega_1(t)$ and $\omega_2(t)$ are the angular velocities of the two beams respectively, $t=0$ is the time when the two beams begin to move around the solenoid, and $t=T$ is the time when the two beams overlap and re-interfere. We have the relation $\int_0^{T}{(\omega_1(t)+\omega_2(t))dt}=2\pi$. 

Here it is worth noting that $\omega_k(t)$ ($k$=1,2) should be determined by the motion of the electron under the influence of the magnetic flux, not by the motion of the free electron. 
As we will see later, 
due to the existence of the induced electric field, one beam will be accelerated and the other beam will be decelerated, and thus the overlapping region will be in general different from the overlapping region for the static case, although the meeting time $T$ are the same for both cases.  

We need to calculate the angular velocity of each electron beam. 
When $\Phi(t)$ varies with time, the motion of the electron will be  changed by the induced electric field. 
For a time-dependent magnetic flux $\Phi(t)$, 
the induced electric field at radius $r$ is: 
\begin{equation}
    \mathbf{E} = E_{\theta}\hat{\bm{\theta}} = -\frac{1}{2\pi r} \frac{d\Phi}{dt}\hat{\bm{\theta}}.\label{E}
\end{equation}
This field will exert a force on the electron, changing its angular velocity. 
The angular momentum equation is:
\begin{equation}
\frac{d}{dt} \big( m r^2(t) \, \omega(t) \big) = - \frac{e}{2\pi} \frac{d\Phi(t)}{dt}.
\label{eq:angular_general}
\end{equation}
Express the angular velocities using the same initial angular momentum $L_0 = m r^2(0)\, \omega(0)$:
\begin{align}
\omega_1(t) &= \frac{L_0 - \frac{e}{2\pi} (\Phi(t) - \Phi(0))}{m r_1^2(t)}, \label{eq:omega_1} \\
\omega_2(t) &= \frac{L_0 + \frac{e}{2\pi} (\Phi(t) - \Phi(0))}{m r_2^2(t)}. \label{eq:omega_2}
\end{align}
Here, the minus sign for path 1 and plus sign for path 2 reflect that the induced tangential electric field accelerates one path clockwise and the other counterclockwise.

Substituting the expressions for $\omega_1(t)$ and $\omega_2(t)$ in (\ref{AB}) gives the AB phase shift: 
\begin{align}
\Delta \phi_{\rm AB} &= \frac{e}{2 \pi m} \int_0^T \Phi(t) \left[
\frac{L_0 - \frac{e}{2\pi} (\Phi(t) - \Phi(0))}{r_1^2(t)}
+ \frac{L_0 + \frac{e}{2\pi} (\Phi(t) - \Phi(0))}{r_2^2(t)}
\right] dt \nonumber \\
&= \frac{e}{2 \pi m} \int_0^T \Phi(t) \left[L_0 \left( \frac{1}{r_1^2(t)} + \frac{1}{r_2^2(t)} \right) - \frac{e}{2\pi} (\Phi(t) - \Phi(0)) \left( \frac{1}{r_1^2(t)} - \frac{1}{r_2^2(t)} \right)\right] dt. \label{AAB}
\end{align}
Similarly, the kinetic phase shift is
\begin{align}
\Delta \phi_{\rm kin} &= \frac{1}{2 m} \int_0^T 
\Bigg[\frac{\big( L_0 - \frac{e}{2\pi} (\Phi(t) - \Phi(0)) \big)^2}{r_1^2(t)} - \frac{\big( L_0 + \frac{e}{2\pi} (\Phi(t) - \Phi(0)) \big)^2}{r_2^2(t)}\Bigg] dt \nonumber \\
&= \frac{1}{2m} \int_0^T \Bigg[L_0^2 \Big( \frac{1}{r_1^2(t)} - \frac{1}{r_2^2(t)} \Big) - \frac{e L_0}{\pi} (\Phi(t) - \Phi(0)) \Big( \frac{1}{r_1^2(t)} + \frac{1}{r_2^2(t)} \Big) \nonumber \\
&\quad + \left( \frac{e}{2\pi} \right)^2 (\Phi(t) - \Phi(0))^2 \Big( \frac{1}{r_1^2(t)} - \frac{1}{r_2^2(t)} \Big)\Bigg] dt.
\end{align}
Adding the kinetic and AB contributions, the total phase shift is
\begin{align}
\Delta \phi_{\rm tot} &= \frac{L_0 e\Phi(0)}{2 \pi m} \int_0^T \left( \frac{1}{r_1^2(t)} + \frac{1}{r_2^2(t)} \right)dt \nonumber \\
&\quad + \frac{1}{2 m} \int_0^T 
\Big( \frac{1}{r_1^2(t)} - \frac{1}{r_2^2(t)} \Big) \left[ L_0^2 - \frac{e^2}{4\pi^2} (\Phi^2(t) - \Phi^2(0)) \right] dt.
\end{align}

Suppose the electron is constrained to a fixed circular path with radius $R$ (e.g., via an external force without affecting tangential dynamics), which will simplify the above results greatly. Then the angular velocities of the two electron beams are
\begin{align}
    \omega_1(t) &= \omega_1(0) - \frac{e}{2\pi m R^2} (\Phi(t)-\Phi(0)), \label{o1}\\
    \omega_2(t) &= \omega_2(0) + \frac{e}{2\pi m R^2} (\Phi(t)-\Phi(0)).\label{o2}
\end{align}
Substituting these two formulae in (\ref{AB}) we obtain the AB phase shift: 
\begin{equation}
    \Delta \phi_{\rm AB} = \frac{e}{2\pi} \int_0^{T} \Phi(t) (\omega_1(0)+\omega_2(0)) dt = \frac{1}{T}  \int_0^{T} e\Phi(t)dt. \label{TDAB}
\end{equation}
Note that $\int_0^{T}{(\omega_1(0)+\omega_2(0))dt}=2\pi$. 
When $\Phi(t)=\Phi_0$, this result reduces to the usual result for the static case $\Delta \phi_{\rm AB} = e\Phi_0$. The kinetic phase shift $\Delta \phi_{\rm kin}$ is
\begin{align}
\Delta \phi_{\rm kin} &= \int_0^T \frac{1}{2}m\big(v_1^2-v_2^2\big)dt 
= \int_0^T \frac{1}{2}mR^2\big(\omega_1(t)^2-\omega_2(t)^2\big)dt \nonumber \\ 
&= e\Phi(0)-\frac{1}{T}\int_0^Te\Phi(t)dt.
\end{align} 
by using (\ref{o1}) and (\ref{o2}). 
Then the total phase shift will be 
\begin{equation}
\Delta \phi_{\rm tot} = \Delta \phi_{\rm AB} + \Delta \phi_{\rm kin} = e\Phi(0).\label{TDT}
\end{equation}
However, independence of \(\Delta \phi_{\rm tot}\) from the time-dependent flux \(\Phi(t)\) is a special feature of circular paths. For general, non-circular paths, the radii $r_k(t)$ ($k=1,2$) vary with time. The angular velocities \(\omega_k(t)\) must be computed using \eqref{eq:omega_1} and \eqref{eq:omega_2}, and the integrals for the kinetic and AB phases generally do not cancel. Therefore, the total phase difference explicitly depends on the full time-dependent flux \(\Phi(t)\) and the path shapes. 
 
Two important points emerge from the above analysis of the time-dependent AB effect. 
First, for constant flux $\Phi(t) = \Phi_0$, the angular velocities satisfy
\begin{equation}
m r_k^2(t) \, \omega_k(t) = L_0 = \text{constant}, \quad k=1,2.
\end{equation}
Then, the AB phase shift (\ref{AAB}) reduces exactly to
\begin{equation}
\Delta \phi_\text{AB} = \frac{e}{2\pi} \int_0^T \Phi_0 \big(\omega_1(t) + \omega_2(t)\big) dt = e \Phi_0.
\end{equation}
independent of the shapes of the paths. This demonstrates that the path independence of the AB phase is a direct consequence of the constant flux and the topological nature of the effect. For time-dependent flux, the integral explicitly depends on $\Phi(t)$ and the paths $r_k(t)$, and the simple path independence no longer holds.
Second, it is crucial to note that Stokes' theorem constrains only the AB phase, which comes from the line integral of the vector potential around the loop,
\begin{equation}
\oint \mathbf{A} \cdot d\mathbf{r} = \int_S \mathbf{B} \cdot d\mathbf{S} = \Phi,
\end{equation}
and is independent of the particle's velocity along the paths. The kinetic phase, in contrast, depends on the detailed time evolution of the particle's speed and trajectory, and is not constrained by Stokes' theorem.  
Therefore, while the AB phase reflects the topological flux through the loop, the total phase generally depends on both the flux history $\Phi(t)$ and the shape of the paths via the kinetic contribution. Only in the special case of constant flux and symmetric circular paths does the total phase reduce simply to $\Delta \phi_\text{tot} = e \Phi_0$.

\section{Consistency with Maxwell’s Equations}

The phase shift derivation in Section 2 employs the magnetic vector potential:
\begin{equation}
    \mathbf{A}(r, t) = \frac{\Phi(t)}{2\pi r} \hat{\boldsymbol{\theta}}, \quad A_0 = 0 \quad (r > a), \label{A_appox}
\end{equation}
where \(\Phi(t)\) is the time-dependent magnetic flux within an infinite solenoid of radius \(a\). This gauge choice implies \(\mathbf{B} = \nabla \times \mathbf{A} = 0\) outside the solenoid (\(r > a\)), suggesting no external magnetic field. However, as noted earlier, this assumption holds only within a quasistatic approximation, since a time-varying \(\Phi(t)\) induces an electric field (\ref{E}), which, per Maxwell’s equations, generates a secondary magnetic field outside the solenoid. Here, we evaluate the validity of this approximation for general time dependencies and assess its consistency with Maxwell’s equations by drawing on rigorous analyses given in \cite{Abbott1985,Templin1995, Zangwill2013, Parker2024}. 

The quasistatic approximation assumes that radiative effects from the time-varying magnetic flux \(\Phi(t)\) are negligible, permitting the simplification that \(\mathbf{B} \approx 0\) outside the solenoid. This requires that \(\Phi(t)\) varies slowly compared to the light-travel times across the solenoid (\( a/c \)) and the observation distance (\( r/c \)), where \(c\) is the speed of light.\footnote{This paper uses units where $\hbar = c = 1$, but retains $c$ in this Section for a clearer explanation.}
The characteristic time scale of variation, \(\tau\), is defined as the duration over which \(\Phi(t)\) undergoes significant change (e.g., the rise time or period for oscillatory fluxes). The approximation is valid when \(\tau \gg a/c\) and \(\tau \gg r/c\), ensuring that electromagnetic wave propagation effects can be ignored. 

For a solenoid with a sinusoidal surface current \(I(t) = I_0 \cos(\omega t)\), 
the exact vector potential outside the solenoid (\(r > a\)) is given by \cite{Abbott1985, Zangwill2013}:
\begin{equation}
    \mathbf{A}(r, t) = \frac{\Phi(0)}{2 a} J_1(k a) \left[ J_1(k r) \sin(\omega t) - Y_1(k r) \cos(\omega t) \right] \hat{\boldsymbol{\theta}}, \label{A_exact}
\end{equation}
where \(\Phi(0) = \frac{4 \pi^2 n I_0 a^2}{c}\) is the corresponding flux amplitude (\(n\) is the number of turns per unit length), 
\(k = \omega / c\), \(J_1\) and \(Y_1\) are Bessel functions of the first and second kind, respectively. The corresponding electric and magnetic fields are:
\begin{align}
    \mathbf{E}(r, t) &= -\frac{\Phi(0) \omega}{2 a} J_1(k a) \left[ J_1(k r) \cos(\omega t) + Y_1(k r) \sin(\omega t) \right] \hat{\boldsymbol{\theta}}, \label{E_exact} \\
    \mathbf{B}(r, t) &= \frac{\Phi(0) \omega}{2 a c} J_1(k a) \left[ J_0(k r) \sin(\omega t) - Y_0(k r) \cos(\omega t) \right] \hat{\mathbf{z}}. \label{B_exact}
\end{align}
In the quasistatic limit (\(\omega a / c \ll 1\) and \(\omega r / c \ll 1\)), using small-argument approximations \(J_1(k a) \approx \frac{\omega a}{2 c}\), \(J_1(k r) \approx \frac{\omega r}{2 c}\), \(Y_1(k r) \approx -\frac{2 c}{\pi \omega r}\), the \(Y_1(k r) \cos(\omega t)\) term dominates in equation (\ref{A_exact}):
\begin{equation}
    \mathbf{A}(r, t) \approx \frac{\Phi(0)}{2 a} \cdot \frac{\omega a}{2 c} \cdot \left( \frac{2 c}{\pi \omega r} \cos(\omega t) \right) = \frac{\Phi(0) \cos(\omega t)}{2 \pi r} = \frac{\Phi(t)}{2 \pi r} \hat{\boldsymbol{\theta}}, \label{A_quasistatic}
\end{equation}
matching equation (\ref{A_appox}). It can also be shown that the magnetic field outside the solenoid $\mathbf{B}(r, t)$ is of order \((\omega a / c)^2\) and thus can be ignored in the quasistatic limit (see \cite{Templin1995}). 

It is worth noting that for a linearly varying flux, \(\Phi(t) = \alpha t\), the second derivative of the surface current vanishes (\(\frac{d^2 I}{dt^2} = 0\)), and the magnetic field outside the solenoid is exactly zero, as derived by Abbott and Griffiths \cite{Abbott1985}. In this case, the quasistatic approximation becomes exact, and equation (\ref{A_appox}) is fully consistent with Maxwell’s equations. 

Recently Parker analyzes the cases of arbitrary time-dependent surface current density \(K(t)=nI(t)\) \cite{Parker2024}, confirming the validity of the quasistatic approximation for general cases. Rather than relying on the vector potential, Parker directly solves Maxwell’s equations using the solenoid’s cylindrical symmetry, reducing them to a pair of coupled partial differential equations for the electric and magnetic fields. His approach yields exact solutions via Green’s functions, enabling the computation of fields for any \(K(t)\). 

These analyses underscore that the quasistatic approximation hinges on the condition \(\tau \gg a/c\) and \(\tau \gg r/c\). For periodic fluxes, \(\tau \approx 2\pi / \omega\), making \(\omega a / c \ll 1\) and \(\omega r / c \ll 1\) the operative criterion. When these ratios are small, the gauge choice \(\mathbf{A}(r, t) = \frac{\Phi(t)}{2\pi r} \hat{\boldsymbol{\theta}}\) and the assumption \(\mathbf{B} \approx 0\) for \(r > a\) are well-justified, ensuring consistency with Maxwell’s equations within the quasistatic regime. 
For rapid variations where \(\omega a / c\) or  \(\omega r / c\) approaches or exceeds unity, radiative corrections become significant, necessitating a fully dynamic treatment beyond the scope of this approximation.

\section{Generalization to Nonzero External Magnetic Field}

The analysis in Section 2 assumed the quasistatic approximation where the magnetic field outside the solenoid is negligible. However, for arbitrary time-varying magnetic fluxes, especially with rapid variations, a nonzero magnetic field $\mathbf{B} \neq 0$ may exist outside the solenoid due to radiative effects. This section extends our derivation to this general case, examining how the external magnetic field affects both the AB phase shift and the kinetic phase shift.

For an ideal infinite solenoid with time-dependent surface current, the exact vector potential outside the solenoid ($r > a$, where $a$ is the solenoid radius) can be derived from Maxwell's equations. In cylindrical coordinates, due to symmetry, $\mathbf{A} = A_\theta(r,t) \hat{\boldsymbol{\theta}}$ and $A_0 = 0$ (in the Lorenz gauge for no free charges). 
The magnetic and electric fields are then:
\begin{align}
\mathbf{B} &= \nabla \times \mathbf{A} = \frac{1}{r} \frac{\partial}{\partial r} (r A_\theta) \hat{\mathbf{z}}, \label{B_general} \\
\mathbf{E} &= -\frac{\partial \mathbf{A}}{\partial t} = -\frac{\partial A_\theta}{\partial t} \hat{\boldsymbol{\theta}}. \label{E_general}
\end{align}
The electron motion is governed by the Lorentz force:
\begin{equation}
m \frac{d\mathbf{v}}{dt} = e (\mathbf{E} + \mathbf{v} \times \mathbf{B}).
\end{equation}
In cylindrical coordinates, with $\mathbf{v} = v_r \hat{\mathbf{r}} + v_\theta \hat{\boldsymbol{\theta}}$, the radial and azimuthal components are:
\begin{align}
m \frac{d v_r}{dt} - m r \omega^2 = e \omega r B_z, \label{radial_eq} \\
m \frac{d}{dt} (r^2 \omega) = e r (E_\theta - v_r B_z). \label{angular_eq}
\end{align}
where $v_r = dr/dt$, and $\omega = v_\theta / r$ is the angular velocity.\footnote{If there are any external forces $F_{\rm ext}$ constraining the path (without affecting tangential dynamics), the radial motion equation will be $m \frac{d v_r}{dt} - m r \omega^2 = e \omega r B_z + F_{\rm ext}$.} 

For general non-circular paths, the total phase shift is:
\begin{equation}
\Delta \phi_{\rm tot} = \Delta \phi_{\rm AB} + \Delta \phi_{\rm kin},
\end{equation}
where
\begin{align}
\Delta \phi_{\rm AB} &= e \int_0^T \left[ A_\theta(r_1,t) r_1 \omega_1 + A_\theta(r_2,t) r_2 \omega_2 \right] dt, \label{AB_general} \\
\Delta \phi_{\rm kin} &= \frac{1}{2} m \int_0^T \left[ (v_{r,1}^2 + r_1^2 \omega_1^2) - (v_{r,2}^2 + r_2^2 \omega_2^2) \right] dt. \label{kin_general}
\end{align}
The angular velocities $\omega_k(t)$ and radial velocities $v_{r,k}(t)$ ($k=1,2$) must be determined by solving equations (\ref{radial_eq}) and (\ref{angular_eq}) along each path. 

For circular paths of radius $R$, the above phase shift expressions simplify to analytically tractable forms.
The AB phase shift is: 
\begin{equation}
\Delta \phi_{\rm AB} = e \int_0^T A_\theta(R,t) R (\omega_1(t) + \omega_2(t)) dt,
\end{equation}
From the equation of motion:
\begin{align}
\omega_1(t) &= \omega_1(0) - \frac{e}{m R} [A_\theta(R,t) - A_\theta(R,0)], \\
\omega_2(t) &= \omega_2(0) + \frac{e}{m R} [A_\theta(R,t) - A_\theta(R,0)],
\end{align}
Since $\int_0^T (\omega_1 + \omega_2) dt = 2\pi$, we have $\omega_1(0) + \omega_2(0) = 2\pi / T$. Thus:
\begin{equation}
\Delta \phi_{\rm AB} = \frac{2\pi e R}{T} \int_0^T A_\theta(R,t) dt = \frac{1}{T} \int_0^T e\Phi_{\rm enc}(R,t) dt,
\end{equation}
where $\Phi_{\rm enc}(R,t) = 2\pi R A_\theta(R,t)$. The AB phase shift is the time average of the enclosed flux. 
Note that $\Phi_{\rm enc}(R,t)$ generally differs from the flux inside the solenoid $\Phi(t)$ when $B_z \neq 0$ outside.

The kinetic phase shift is:
\begin{equation}
\Delta \phi_{\rm kin} = \frac{1}{2} m R^2 \int_0^T (\omega_1^2 - \omega_2^2) dt.
\end{equation}
Using the expressions for $\omega_1(t)$ and $\omega_2(t)$ and for symmetric initial conditions $\omega_1(0) = \omega_2(0)$, we obtain:
\begin{equation}
\Delta \phi_{\rm kin} = \frac{2\pi e R}{T} \int_0^T [A_\theta(R,0) - A_\theta(R,t)] dt.
\end{equation}
The total phase shift is:
\begin{equation}
\Delta \phi_{\rm tot} = \Delta \phi_{\rm AB} + \Delta \phi_{\rm kin} = e \Phi_{\rm enc}(R,0),
\end{equation}
where $\Phi_{\rm enc}(R,0) = 2\pi R A_\theta(R,0)$. 
Thus, for circular paths, the total phase shift depends only on the initial enclosed flux, regardless of the time variation of the flux or the presence of $B_z \neq 0$ outside.

This analysis reveals several important features of the generalized AB effect with nonzero external magnetic field:

\begin{enumerate}
\item For circular paths of radius $R$, the total phase shift reduces to $e \Phi_{\rm enc}(R,0)$ in the symmetric case. The external $B$ field does not directly appear in this result because for circular motion with constant $R$, the $v_r B_z$ term in the equation of motion (\ref{angular_eq}) vanishes.

\item For general non-circular paths, the external $B$ field affects both the electron trajectories and the phase accumulation through the $v_r B_z$ term in equation (\ref{angular_eq}). This leads to additional path dependence beyond what is present in the quasistatic case.

\item The AB phase shift $\Delta \phi_{\rm AB}$ remains gauge-invariant and is determined by the line integral of $\mathbf{A}$ around the closed loop. However, in the time-dependent case with nonzero $B$ field outside, the enclosed flux $\Phi_{\rm enc}(r,t)$ is evaluated along the actual paths, which generally differs from the flux inside the solenoid $\Phi(t)$. 

\item The kinetic phase shift $\Delta \phi_{\rm kin}$ contains contributions from both the induced electric field and the external magnetic field, making it gauge-invariant but path-dependent.

\item In the quasistatic limit, the external $B$ field becomes negligible, and we recover the results of Section 2. For rapid flux variations, the full dynamic treatment presented here is necessary.
\end{enumerate}

These results demonstrate that the time-dependent AB effect with nonzero external magnetic field exhibits rich behavior, with the total phase shift depending on both the gauge potential (AB term) and the electromagnetic fields (kinetic term) in a path-dependent manner. This hybrid nature underscores the fundamental interplay between gauge symmetry and physical forces in quantum dynamics.

\section{Implications for Quantum Phase Dynamics}

Our WKB-based derivation of the time-dependent AB phase shift offers new insights into the dynamic generation of quantum phases, addressing longstanding debates and limitations in prior studies \cite{Lee1992, Gaveau2011, SV2013, MS2014, BS2015, AS2016, JZWLD2017, CM2019, Wakamatsu2024, Wakamatsu2025}. Previous analyses have often yielded conflicting results due to incomplete treatments of electron dynamics and induced fields. For instance, Singleton et al. \cite{SV2013, MS2014} concluded that the phase shift remains static, equivalent to the initial flux for an arbitrary time-dependent flux. However, their derivations based on the 4-dimensional Stokes’ theorem are problematic \cite{Wakamatsu2024, Wakamatsu2025}. Conversely, other works \cite{JZWLD2017, CM2019} proposed an instantaneous phase shift $e \Phi(t)$, which fails to account for the temporal accumulation of the phase over the electron's travel time, during which the flux evolves. Lee et al.'s approach \cite{Lee1992} is closest to ours, incorporating dynamic motion, but remains complex and lacks generality for arbitrary paths and flux profiles (e.g. ignoring the kinetic phase shift). Moreover, earlier studies inadequately addressed the quasistatic approximation's validity; for oscillatory currents, the period must exceed the electron's traversal time $T$ to neglect radiative effects, limiting applicability to slow variations (cf. \cite{Lee1992}).

The key innovation in our analysis is the WKB decomposition of the total phase into AB ($\Delta \phi_{\rm AB}$) and kinetic ($\Delta \phi_{\rm kin}$) components, explicitly incorporating the electron's trajectory under the induced electric field. This reveals that, within the quasistatic regime, the AB phase shift for circular paths is proportional to the time-averaged enclosed magnetic flux, $\Delta \phi_{\rm AB} = \frac{1}{T} \int_0^T e \Phi(t) , dt$, while the kinetic contribution counterbalances it to yield a total phase $\Delta \phi_{\rm tot} = e \Phi(0)$. For non-circular paths, the phase shift becomes path-dependent, with both components influenced by the flux history $\Phi(t)$ and radii $r_k(t)$, highlighting the effect's sensitivity to geometry in dynamic scenarios. Notably, for constant flux $\Phi(t) = \Phi_0$, the result recovers the path-independent static AB phase $e \Phi_0$, underscoring the topological robustness in time-independent cases.

The generalization to nonzero external magnetic fields, as detailed in Section 4, extends these results by considering the enclosed flux through the actual electron paths, \(\Phi_{\rm enc}(t)\), rather than solely the solenoid’s flux \(\Phi(t)\). For circular paths, the AB phase shift is the time average of \(\Phi_{\rm enc}(R,t)\), with the total phase depending only on the initial enclosed flux \(e \Phi_{\rm enc}(R,0)\). For general non-circular paths, the external $B$ field affects trajectories and phase accumulation through the Lorentz force, leading to additional path dependence, with the kinetic phase including contributions from both induced electric and external magnetic fields. This reduces to the quasistatic results when external fields are negligible. 

This generalized AB effect is inherently hybrid: the electron encounters an induced electromagnetic field, so the phase arises from both gauge-dependent potentials (AB term) and gauge-invariant field forces (kinetic term). The kinetic phase, stemming from the solenoid's time-varying current, is gauge-invariant even for open paths and can be mitigated by external forces to nullify net acceleration. For example, consider a flux profile active only during a fraction of the traversal, e.g., constant $\Phi_0$ from $T/4$ to $3T/4$ with linear ramps ensuring zero external magnetic field \cite{Abbott1985, Zangwill2013, Parker2024}. Our result predicts $\Delta \phi_{\rm AB} \approx \frac{1}{2} e \Phi_0$ for circular paths, predominantly from the vector potential when field effects are nullified by external forces.

These results may have profound implications for the AB effect's foundational interpretation. The time-averaged phase shift supports a continuous, local accumulation during the electron's traversal, rather than an instantaneous manifestation at interference. 
In the intermittent flux example, 
when there is a constant magnetic flux inside the solenoid during the time interval $[T/4, 3T/4]$, no gauge-invariant quantities of the electron are affected by the magnetic flux inside the solenoid. While when the electron beams overlap, there is no magnetic flux inside the solenoid anymore, and thus the motion of the electron is not affected by the electromagnetic field either. 
This strongly favors a continuous, local potential explanation of the AB effect and disfavors a discontinuous, nonlocal field explanation of the AB effect \cite{Gao2025}. 

\section{Summary and Future Directions}

This study derives the AB phase shift for a time-dependent magnetic vector potential using the WKB method, providing a robust framework for the generalized AB effect. For circular paths, the AB phase shift is proportional to the time-averaged enclosed magnetic flux, while the total phase shift, including the kinetic contribution, depends only on the initial enclosed flux. For non-circular paths, the total phase shift depends on both the flux history and the path geometries, revealing the effect’s dependence on electron trajectories and induced electromagnetic fields. This hybrid nature, blending gauge-dependent potentials and gauge-invariant field effects, resolves discrepancies in prior studies and supports a continuous, local potential interpretation of the AB effect \cite{Gao2025}. The derivation’s consistency is ensured through verification with Maxwell’s equations.

To date, experimental tests of the time-dependent AB effect, such as Ageev et al. \cite{Ageev2000}, have not observed a phase shift, possibly due to insufficient sensitivity or non-ideal conditions. An earlier experiment by Marton et al. \cite{Marton1954} inadvertently probed this effect but also reported no shift. These null results highlight the need for high-precision experiments to test the predicted phase shift \cite{Werner1960, Macdougall2015}. Future work should focus on designing experiments with controlled varying magnetic fluxes and improved detection techniques. Additionally, exploring relativistic effects or extensions to other quantum systems with time-dependent potentials could further elucidate the dynamics of phase accumulation. 
The confirmation of the generalized AB effect would not only deepen our understanding of quantum mechanics and gauge theories but also offer new theoretical insights and potential applications in quantum technologies.

\bibliographystyle{plain}

\end{document}